\documentclass[sigconf,screen]{acmart}

\usepackage{amsmath,amsfonts}
\usepackage{algorithmic}
\usepackage{graphicx}
\usepackage{textcomp}
\usepackage{makecell}
\usepackage{xcolor}
\usepackage{xurl}
\usepackage{url}
\def\BibTeX{{\rm B\kern-.05em{\sc i\kern-.025em b}\kern-.08em
    T\kern-.1667em\lower.7ex\hbox{E}\kern-.125emX}}

\setcopyright{none}
\acmDOI{}
\acmISBN{}
\acmPrice{}
\acmYear{2026}

\acmBooktitle{Companion Proceedings of the 34th ACM Symposium on the Foundations of Software Engineering (FSE '26), June 5--9, 2026, Montreal, Canada}

\acmConference[FSE '26]{Companion Proceedings of the 34th ACM Symposium on the Foundations of Software Engineering (FSE '26)}{June 5--9, 2026}{Montreal, Canada}

\begin{document}

\title{Attesting LLM Pipelines: Enforcing Verifiable Training and Release Claims}

\author{Zhuoran Tan}
\affiliation{%
  \institution{University of Glasgow}
  \country{UK}
}

\author{Jeremy Singer}
\affiliation{%
  \institution{University of Glasgow}
  \country{UK}
}

\author{Christos Anagnostopoulos}
\affiliation{%
  \institution{University of Glasgow}
  \country{UK}
}

\begin{abstract}
Modern Large Language Model(LLM) systems are assembled from third-party artifacts—pre-trained weights, fine-tuning adapters, datasets, dependency packages, and container images—fetched through automated pipelines. This speed comes with supply-chain risks: compromised dependencies, malicious hub artifacts and unsafe deserialization, forged provenance, and backdoored models. A core gap is that training and release claims (e.g., data and code lineage, build environment, and security scanning results) are rarely cryptographically bound to the artifacts they describe, making enforcement inconsistent across teams and stages.
We propose an attestation-aware promotion gate: before an artifact is admitted into trusted environments (training, fine-tuning, deployment), the gate verifies claim evidence, enforces safe-loading and static scanning policies, and applies secure-by-default deployment constraints. When organizations operate runtime security tooling, the same gate can optionally ingest standardized dynamic signals via plug-ins to reduce uncertainty for high-risk artifacts. We outline a practical claims-to-controls mapping and an evaluation blueprint using representative supply-chain scenarios and operational metrics (coverage, and decisions), charting a path toward a full research paper.
\end{abstract}



\maketitle

\section{Introduction \& Position Statement}

Arifical Intelligence (AI) systems are increasingly assembled from third-party artifacts---pre-trained models, fine-tuning adapters, datasets, and dependency packages---that are fetched, composed, and deployed through automated pipelines \cite{openssf_mlsecops_2025}. This ``AI supply chain'' accelerates delivery but expands the attack surface: adversaries can compromise dependency ecosystems~\cite{itach2023pytorch}, inject malicious code into model hub artifacts and unsafe serialization paths \cite{10.1145/3691620.3695271}, or distribute backdoored models whose malicious behavior is activated by rare triggers \cite{verma2025exploiting}. Unlike traditional software supply chains, AI artifacts often include opaque weight tensors and auxiliary code (custom layers, converters, loaders), while the provenance of training and fine-tuning is frequently incomplete or unverifiable \cite{Bennet_2024,AISecurityRiskAssessment}.

Current practice adopts methods from software security (Software Composition Analysis, Software Bills of Material, CI/CD hardening), however, critical gaps remain. First, provenance and transparency are weaker: model origins and training context are rarely verified bound to artifacts \cite{Bennet_2024}. Second, model hubs blend weights with code, enabling malicious loaders or unsafe deserialization to execute during import and model loading \cite{10.1145/3691620.3695271}. Third, integrity checks alone cannot rule out behavior-level compromise such as backdoors \cite{verma2025exploiting}. Therefore, prior work calls for more rigorous verification of AI artifacts before deployment \cite{10.1145/3704724}.

\textbf{Our goal} is to enable attestable LLM pipelines by making training and release claims verifiable and enforceable: before an artifact is promoted into a trusted environment (training, fine-tuning, deployment), an attestation-aware gate validates claim evidence and applies policy-based admission decisions.
The contribution is:
\begin{itemize}
    \item An attestation-aware promotion gate that enforces verifiable training and release claims before LLM artifacts are admitted into trusted environments.
    \item A concrete threat model and a claims-to-evidence and threat-to-control mapping for key LLM supply-chain risks (compromised dependencies, malicious hub artifacts, provenance forgery, and backdoored models). Our mapping is informed by a survey of stage-wise defenses from both industry and academia, but centers on a unified, auditable, and verifiable end-to-end supply-chain gate.
    \item We outline an evaluation blueprint and a path to a paper, including case studies and operational metrics, and discuss when optional dynamic plug-ins can reduce uncertainty.
\end{itemize}

\section{Claim Taxonomy \& Threat Model}

\begin{figure*}[htbp]
    \centering
    \includegraphics[width=0.85\textwidth]{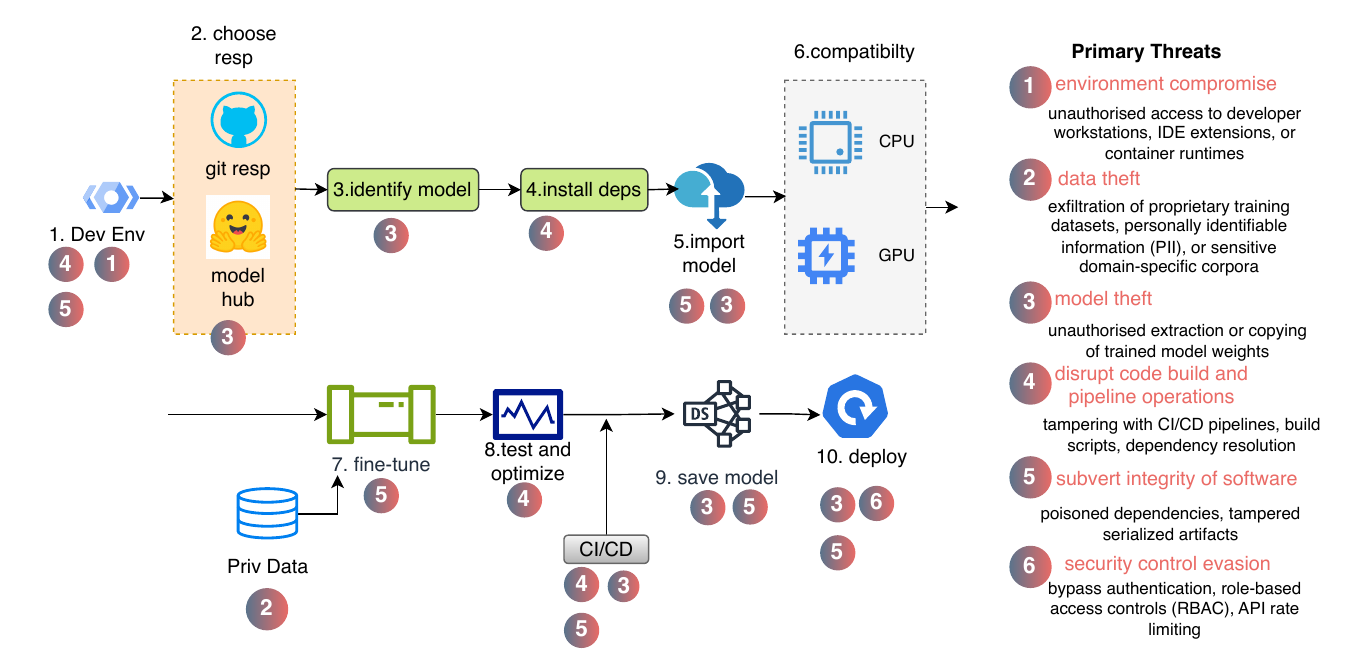}
    \caption{Threat Model of AI Supply Chain}
    \label{fig: threat model}
\end{figure*}

\subsection{Training Claims}
Training claims describe how an artifact was produced and which inputs influenced its behavior. We focus on claims that can be recorded and bound to a training run and artifact digest:
\begin{enumerate}
    \item data lineage: dataset identifiers, versions/digests, and sampling policies;
    \item code lineage: training scripts, configuration, and commits;
    \item dependency \& environment snapshot: lockfiles, container/base image digest, and critical library versions;
    \item hyperparameter summary: optimizer, learning rate schedule, batch size, and random seeds (hashed);
    \item compute context: accelerator type and driver/toolchain versions (hashed);
    \item run metadata: start/end times, training logs checksum, and output artifact digests.
\end{enumerate}

Claims are issued as in-toto~\footnote{https://in-toto.io/} attestation predicates, signed via Sigstore~\footnote{https://www.sigstore.dev/} and bound to the artifact's content digest. This binds the claims to the exact bytes being deployed, making them verifiable and non-transferable: consumers can cryptographically confirm the claims apply to the specific artifact they pulled, avoiding tag/filename ambiguity and preventing artifact substitution \cite{slsa_attestation_model}.

\subsection{Release Claims}

Release claims constrain how an artifact can be safely consumed and deployed. We emphasize enforceable claims that map directly to admission controls:
\begin{enumerate}
    \item artifact identify: name/version plus content digest; optional signing identity;
    \item format guarantees: e.g., safe tensor formats and explicit disallow-list (e.g., pickle) for unsafe deserialization;
    \item embedded-code declaration: whether custom code, converters, or loaders are present;
    \item static scan results: malware/serialization/package scan summaries and tool versions;
    \item evaluation/security summary: minimal checks (sanity tests, regression/smoke tests) and their hashes;
    \item deployment requirements: required permissions (egress, filesystem (FS), GPU access) expressed as a policy contract.
\end{enumerate}

Release claims are serialized as a CycloneDX 1.6 MLBOM~\footnote{https://cyclonedx.org/capabilities/mlbom/} bundled with evidence (scan reports, Sigstore signatures) and validated by the promotion gate prior to deployment.

\subsection{Threat Model}

We consider a typical workflow in which a pre-trained LLM is deployed and further fine-tuned to support diverse downstream services. In this pipeline, adversaries may introduce or modify components at multiple stages (Figure~\ref{fig: threat model})~\cite{10.1145/3704724,openssf_mlsecops_2025}. These threats include compromised dependencies, malicious artifacts from model hubs (e.g., unsafe loaders or serialization mechanisms), and forged or missing provenance metadata.
We assume the admission environment is capable of enforcing policy checks and maintaining audit logs. However, we do not assume the universal availability or reliability of signed provenance~\cite{slsa_attestation_model}. In the absence of trustworthy provenance, the admission gate adopts conservative controls, including safe-loading mechanisms and static analysis, and places artifacts into quarantine when supporting evidence is insufficient.
We note that behavior-level threats, such as subtle model backdoors~\cite{zeng2025clibe}, are not fully detectable through static analysis alone. Addressing such risks requires complementary measures, including optional lightweight validation signals and risk-based triage.

\begin{figure*}[htbp]
    \centering
    \includegraphics[width=0.9\textwidth]{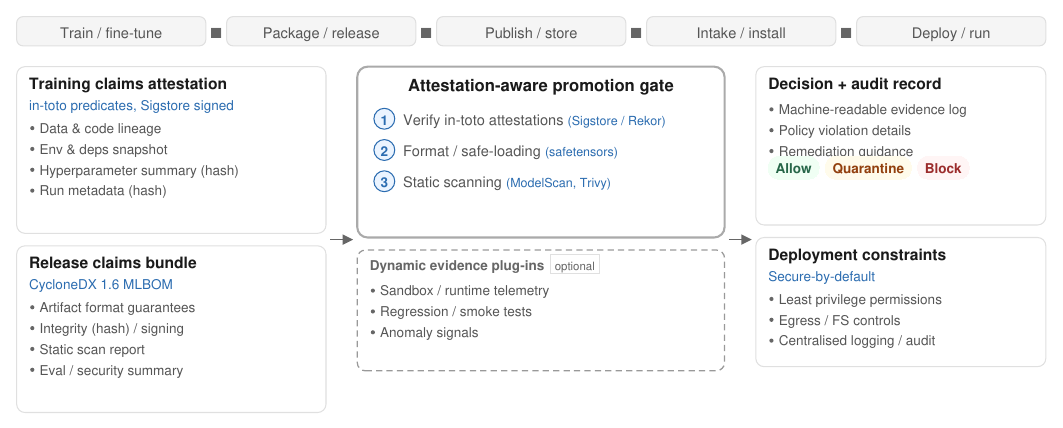}
    \caption{Claims $\rightarrow$ Evidence $\rightarrow$ Promotion Gate $\rightarrow$ Decision (Attesting LLM Pipelines)}
    \label{fig:llm attest framework}
\end{figure*}

\section{Attestation-aware Promotion Gate}

The promotion gate is an admission-control point that enforces claims before artifacts are promoted into trusted environments (training, fine-tuning, deployment). As summarized in Figure~\ref{fig:llm attest framework}, the gate ingests an artifact together with its training and release claim bundles, validates evidence (when available), and applies policy-based decisions: allow, quarantine, or block—always producing an auditable, machine-readable record. Concretely, the gate (i) performs schema/format checks and safe-loading enforcement, (ii) verifies Sigstore signatures and in-toto provenance attestations when present, (iii) runs static scanning over packages and serialization surfaces, and (iv) optionally consumes standardized dynamic signals from existing runtime tooling for high-risk artifacts.

\subsection{Deployment Constraints}
Given that prevention is challenging, we enforce safe-by-default deployment constraints: least-privilege permissions, controlled network egress, restricted files-system access, and centralized logging consistent with industry guidance \cite{AISecurityRiskAssessment}. These constraints reduce blast radius and make post-incident investigation faster and more reliable by preserving high-quality audit trails.

\subsection{Pluggable Dynamic Evidence}
Dynamic execution monitoring can strengthen detection of malicious hub artifacts and compromised dependencies, but implementing a robust sandbox is engineering-heavy. Instead, the gate treats dynamic analysis as an \emph{optional evidence source}: deployments may connect existing tools (e.g., hardened container runtimes or runtime security monitors) to provide standardized signals such as outbound network attempts, suspicious process spawning, and unexpected file writes \cite{10.1145/3691620.3695262}. When such signals are unavailable, the gate still operates using provenance, static scanning, and safe-loading policies; dynamic plug-ins primarily improve coverage and reduce uncertainty for high-risk artifacts.

\section{Walkthrough Case}

Malicious hub artifact with unsafe deserialization. A team pulls a third-party checkpoint and tokenizer from a public hub for fine-tuning. The artifact arrives with incomplete release claims (no signing identity) and is packaged in a format that permits code execution during load. The gate first enforces format policies (e.g., disallow unsafe deserialization) and runs static scanning to extract and inspect embedded scripts/metadata. Because claim evidence is insufficient, the gate quarantines the artifact and emits an audit record capturing missing claims, scan findings, and required remediation (e.g., repackage into an allowed format, or provide signing/provenance). If a dynamic plug-in is available, the organization may execute a lightweight sandbox load to collect signals (unexpected process spawning or outbound network attempts); high-risk signals escalate the decision to block. This walkthrough illustrates how claims drive enforceable admission decisions and how quarantine supports human-in-the-loop triage.

\section{Evaluation Blueprint \& Path to Full Paper}

\begin{table*}[htbp]
\centering
\footnotesize
\caption{Stage-wise Components and Reference Practices in the LLM Supply Chain}
\begin{tabular}{l p{6cm} p{9cm}}
\hline
\textbf{Stage}     & \textbf{Components}         & \textbf{Practice}                                                        \\ \hline
Development        & IDE, Dependencies, Packages &   Dependency-Check\footnote{https://github.com/dependency-check/DependencyCheck}, Trivy \cite{trivy}                                                                     \\ \hline
Model Storage      &   Architecture, Weights, Actor                          &Safetensors\footnote{https://huggingface.co/docs/safetensors/index} Format, SavedModel~\cite{10.1145/3704724}, GGUF and ONNX~\cite{10.1145/3691620.3695271}, CFEs~\cite{sidhpurwala2024buildingtrustfoundationssecurity}, ModelScan \cite{modelscan} \\ \hline
Model Download     &   Model Hub, Models, Script     & Integrity Check~\cite{10.1145/3704724}, MALHUG~\cite{10.1145/3691620.3695271}, AI Bill of Materials (AIBOMs)~\cite{10.1145/3643662.3643957,sidhpurwala2024buildingtrustfoundationssecurity}, ModelScan \cite{modelscan}                                            \\ \hline
Dependency Install &  dependent libraries    &  MDVul (fusion path)\cite{ZHEQU2026103475}, Dependency-Check\footnote{https://github.com/dependency-check/DependencyCheck}                                                                \\ \hline
Fine Tuning        & Adapters (Lora, Peft)                    &  Static Code Analysis (SAST)~\cite{10.1145/3691620.3695262},  Dependency-Check\footnote{https://github.com/dependency-check/DependencyCheck}                                                                     \\ \hline
CI/CD              &  Build System, Testing, Deployment Platform, Source Code                           &  Supply-chain Levels for Software Artifacts (SLSA)~\cite{slsa}, Sigstore~\cite{10.1145/3548606.3560596}, Trivy \cite{trivy}, ModelScan \cite{modelscan} \\ \hline
Deployment         & Model, Platform, Interface  & Rigorous Validation Process (simulate attacks, test robustness)~\cite{10.1145/3704724}           \\ \hline
Usage              & Input, Output               & Automated Content-Filtering, Anomaly Detection~\cite{10.1145/3704724}                            \\ \hline
\end{tabular}
\end{table*}

We outline an evaluation blueprint for the promotion gate along three practical dimensions:
(i) \emph{threat coverage}---which supply-chain attack surfaces are mitigated by enforceable claims and controls;
(ii) \emph{operational overhead}---latency/throughput impact on artifact intake and deployment;
and (iii) \emph{triage usability}---whether the produced evidence is sufficient to support fast, auditable decisions.

\subsection{Coverage via Case Studies}
We use representative scenarios spanning common LLM supply-chain threats:
(1) dependency confusion-style package compromise, reproduced following the PyTorch attack vector documented in \cite{itach2023pytorch} against a controlled package registry;
(2) malicious model-hub artifacts and unsafe deserialization paths, evaluated against the malicious artifact corpus collected from Hugging Face by MALHUG~\cite{10.1145/3691620.3695271};
and (3) backdoored-model risk as a high-impact but difficult-to-verify class, using backdoor-injected checkpoints following the methodology of \cite{verma2025exploiting} to test quarantine and triage decisions under insufficient provenance.
For each scenario, we record: (a) which checks and policies trigger (e.g., format/safe-loading enforcement, static scanning, provenance/signature validation, and optional dynamic signals), (b) the evidence emitted in the audit record (missing/failed claims, scan findings, policy violations), and (c) the resulting admission outcome (\texttt{allow}/\texttt{quarantine}/\texttt{block}). Coverage is reported as the fraction of threat-scenario steps for which at least one gate policy fires.

\subsection{Operational Overhead and Triage Cost}
We plan to measure gate latency per artifact and per stage (hashing, scanning, provenance/signature checks), and quantify throughput under parallel intake. Beyond automated cost, we quantify the human-in-the-loop burden introduced by \texttt{quarantine}, by measuring the proportion of quarantine records that contain actionable remediation guidance, policy-exception frequency, and false positives on benign artifacts from a curated clean-artifact baseline. These measurements characterize the practical trade-off between strictness and availability. 

\subsection{Scope and Path to a Full Paper}
Our design improves practical resilience but does not claim completeness against adaptive adversaries. Provenance assurances depend on ecosystem adoption, and behavior-level risks (e.g., subtle backdoors) may require task-specific validation. A full paper will report empirical results on the above corpora and study risk-based policy tuning by artifact criticality.

\section{Related Work}

\subsection{Provenance and AI Artifact Inventory}

Prior work spans traditional software supply-chain security and emerging AI-specific artifact management, with notable differences in scope. 
AIBOM~\cite{openssf_mlsecops_2025} extends software inventory concepts to AI by capturing heterogeneous artifacts such as datasets, models, and training processes \cite{Bennet_2024}, reflecting the data-dependent and non-deterministic nature of AI pipelines. 
In contrast, frameworks like SLSA~\cite{slsa,10.1145/3548606.3560596} and Sigstore focus on software integrity through reproducible builds and cryptographic signing, but largely overlook AI-specific challenges such as data lineage, model evolution, and runtime behavior. 
Industry guidance highlights lifecycle governance and traceability in AI systems \cite{AISecurityRiskAssessment}, yet a unified framework that integrates software supply-chain guarantees with AI-specific provenance remains an open challenge.

\subsection{Scanning and Validation of ML Artifacts}
Model and dataset hubs introduce risks because artifacts can include code and unsafe serialization. MALHUG proposes an end-to-end pipeline for detecting malicious code poisoning targeting model/data hubs via script extraction, deserialization analysis, and semantic checks \cite{10.1145/3691620.3695271}. Practical scanners and safer formats aim to reduce load-time code execution risks \cite{modelscan,10.1145/3704724}. Dynamic analysis for package ecosystems shows that sandbox-like execution and behavior tracing can be valuable under obfuscation \cite{10.1145/3691620.3695262}; in our system, such dynamic evidence is treated as an \emph{optional plug-in} rather than a required component.

\section{Conclusion}
We argue that verifiable training and release claims should be enforced through an attestation-aware promotion gate that produces auditable admission decisions. This position paper outlines the threat model, claim taxonomy, and evaluation blueprint toward a full empirical study.


\end{document}